\documentclass[numberedheadings]{aipproc}
\layoutstyle{6x9}

\begin{document}
\title{Neutrinos in cosmology, with some significant digressions}
\author{Raymond R. Volkas}{address={School of Physics, Research Centre
for High Energy Physics,
The University of Melbourne, Victoria 3010 Australia}}

\begin{abstract}
Neutrinos play prominent roles in both particle physics and cosmology.
In this talk, I will cover two broad topics.
The first will be possible origins for neutrino masses and mixings and the
implications of this physics for cosmology. Some non-cosmological digressions
on the flavour problem in general will be made. The second topic
will be Big Bang Nucleosynthesis (BBN) and bounds on active-sterile
neutrino mixing.
\end{abstract}

\maketitle

\section{Survey of Neutrino Cosmology}

Two important issues come to mind when contemplating neutrinos
in cosmology: the possible existence of additional neutrino or 
neutrino-like species, followed by the dynamical implications of neutrino
mass and mixing \cite{dolgov}.

\subsection{Additional neutrino species}

One may classify theoretically possible additional neutrino states into four
categories:
\begin{itemize}
\item {\it Light active.} This option is clearly ruled out by the
measured invisible width of the $Z$ boson: we know that there are
only three active neutrinos with masses less than about $45$ GeV
($\nu_{e,\mu,\tau}$).
\item {\it Light sterile.} The combined solar, atmospheric and
LSND anomalies imply the existence of at least one light sterile
flavour, if one demands that neutrino oscillations account for all
of these data \cite{nuoscdata}. (The reason is simply that the three 
very different $\Delta m^2$ parameters required to furnish the
appropriate oscillation lengths cannot be obtained from just
the three known flavours of light active neutrino.)
\item {\it Heavy active.} Additional active states with masses
greater than $45$ GeV are not precluded by present electroweak
data. We will discuss how such states might arise in extensions
of the standard model (SM) shortly.
\item {\it Heavy sterile.} Various see-saw models of neutrino mass
provide good particle physics 
motivations for this class of particle. In addition
to the standard see-saw mechanism \cite{standardseesaw}, I will also review the
universal \cite{universalseesaw}
and mirror see-saw \cite{mirrorseesaw} frameworks to show that
several different patterns of light and heavy sterile neutrinos can
arise. Perhaps the most important cosmological role proposed
for heavy sterile states is baryogenesis via sphaleron-reprocessed
lepton asymmetries produced either through the decay
or the interactions of the heavy neutral leptons.
\end{itemize}

\subsubsection{The active/sterile distinction \cite{erice}}

By an ``active'' neutrino we mean, of course, one that couples
to the known left-handed weak interaction. The standard
left-handed neutrinos, for example, sit in doublets of electroweak
SU(2)$_L$ with weak-hypercharge $Y = -1$ [normalisation is
such that electric charge $Q = (\tau_{3L} + Y)/2$]. Exotic active
neutrinos are also theoretically possible, perhaps as
components of higher-dimensional weak isospin representations.

The term ``sterile neutrino'' connotes a charge and colour neutral
fermion that is also a singlet of SU(2)$_L$, so that
it does not feel the weak interaction either. This terminology is
appropriate in the context of how experiments are done: the weak force
is the only known way of detecting neutrinos, so any neutrino-like species
immune to this interaction will reveal itself only by its absence!
However, in extensions of the SM it is quite common to find sterile
neutrinos that do in fact feel some as yet hypothetical interaction,
a right-handed weak force for example. 

This motivates a division of sterile neutrinos into two further
categories: ``weakly'' and ``fully'' sterile. Weakly sterile species
are those that feel some hypothetical gauge interaction, while
fully sterile states are those that couple to no gauge force,
known or unknown. This distinction can be important, especially
for theories of sterile neutrino mass and perhaps also for 
cosmology.\footnote{Note that my definition indulges in ``gauge 
interaction chauvinism'' by ignoring spin-0 boson induced
interactions. No slight is intended. Indeed, Higgs-induced
active-sterile mixing is obviously a very important effect.}

Examples of fully sterile neutrinos include the gauge singlets
usually termed ``right-handed neutrinos'' that may be added to
the minimal SM fermions, under the assumption of a gauge force
desert up to the Planck scale. Another related example is the
set of SU(5) singlets one may add to the $5^* \oplus 10$ 
multiplets that compose the standard families in SU(5) grand unified
theories (GUTs). These singlets are fully sterile if SU(5)
is not embedded in a larger group such as SO(10) or E$_6$.

As is well known, fully sterile species allow
gauge invariant bare Majorana mass terms. Since these parameters
are not related to any symmetry breaking vacuum expectation values (VEVs),
they have no natural scale. While many model builders set
these masses to be large as a matter of course, it should be noted
that taking them to zero increases the symmetry of the theory. This means
that small bare Majorana masses for fully sterile neutrinos are
technically natural (i.e.\ preserved under radiative corrections).

Weakly sterile neutrinos are probably more common in SM extensions.
If the Glashow-Weinberg-Salam SU(2)$_L\otimes$U(1)$_Y$ electroweak
model is extended to the left-right symmetric
SU(2)$_L\otimes$SU(2)$_R\otimes$U(1)$_{B-L}$, then the right-handed
neutrinos are in doublets of SU(2)$_R$ together with the
right-handed charged leptons. This kind of right-handed neutrino
couples to the right-sector $W$ and $Z$ bosons so it is weakly
sterile. In many models, the Majorana mass for such a state is
proportional to the right-handed weak isospin symmetry
breaking scale, which is experimentally constrained to be large.
The left-right symmetric model may be embedded in SO(10)
and E$_6$ GUTs. If so, then the (usually heavy) weakly-sterile
neutrinos feel gauge interactions beyond the right-handed weak
force, and they may pick up GUT symmetry-breaking scale masses.
These examples illustrate how heavy sterile neutrinos typically arise
in well-motivated SM extensions.

Mirror matter models \cite{mmm}
provide an interestingly different
class of weakly sterile neutrino. Consider a model with gauge
group $G_{SM}\otimes G_{SM}$, with an exact discrete symmetry
between the sectors. This is usually selected to be a
non-standard parity symmetry so that the 
world is invariant under the full Poincar\'{e}
group, including improper transformations (hence the designation
``mirror matter''). Every ordinary particle has a mirror partner.
The mirror neutrinos are immune to the ordinary weak interaction,
but feel the mirror weak force, so they are weakly sterile. 
However, the exact discrete symmetry ensures that the mirror
neutrinos are light for exactly the same reasons that ordinary
neutrinos are light, whatever those reasons are! This illustrates
how light (weakly) sterile neutrinos can arise in a simple and
well-motivated extension of the SM. The mirror matter model 
provides a qualitatively different outcome for the sterile neutrino
masses compared to most other models because the additional
chiral interactions felt by these states are broken at a
relatively low scale (the electroweak scale in fact).

\subsubsection{Three see-saw scenarios}

I now want to present three case studies of see-saw models of neutrino
mass that can furnish qualitatively different varieties of
additional neutrino-like states, both heavy and light.
I will restrict myself to one-family examples for simplicity,
with the realistic three-family extensions being (mostly)
straightforward generalisations.

The standard see-saw mechanism arises when one adds a gauge-singlet
neutrino $\nu_R$ to a minimal SM family and writes down all
renormalisable gauge-invariant terms \cite{standardseesaw}. The neutrino mass
matrix is contained within the mass term
\begin{equation}
\left( \begin{array}{cc} \overline{\nu}_L & \overline{(\nu_R)^c}
\end{array} \right)
\left( \begin{array}{cc} 0 & m \\ m & M \end{array} \right)
\left( \begin{array}{c} (\nu_L)^c \\ \nu_R \end{array} \right) + H.c.
\end{equation}
where $m$ is the electroweak-scale Dirac mass and $M$ is a bare Majorana
mass. In many obvious SM extensions, $M$ becomes proportional
to a high gauge symemtry breaking scale. If indeed $M \gg m$, then
the eigenvalues have approximate magnitudes $m^2/M$ and $M$, 
with the eigenstates being predominantly $\nu_L$ and $\nu_R$,
respectively. Hence the standard active neutrino $\nu_L$ has
a mass suppressed relative to the electroweak scale by the
small hierarchy parameter $m/M$, while its right-handed
fully sterile partner is very heavy. (In the extensions, the
heavy partner becomes weakly sterile.)

My second case study is the universal see-saw scenario \cite{universalseesaw}. 
One adds
heavy weak-isosinglet vector-like 
partners $U$, $D$, $N$ and $E$ to each of
the standard fermions $u$, $d$, $\nu$ and $e$ and one works
within the left-right symmetric augmentation of the electroweak
sector. The idea is to explain why all fermions (top quark
excluded), not just neutrinos, have small masses relative to
the electroweak scale $\sim 200$ GeV.

Given a Higgs sector containing both SU(2)$_L$ and SU(2)$_R$
doublets but no bidoublets,
the charge $-1/3$ quarks develop a mass matrix as given by
\begin{equation}
\left( \begin{array}{cc} \overline{d}_L & \overline{D}_L
\end{array} \right)
\left( \begin{array}{cc} 0 & m_L \\ m_R & M \end{array} \right)
\left( \begin{array}{c} d_R \\ D_R \end{array} \right) + H.c.
\end{equation}
where $m_L (m_R)$ is proportional to the left-(right-)handed
weak isospin breaking scale, and $M$ is the bare gauge-invariant
Dirac mass for the vector-like $D$ (or it is induced
by a Higgs singlet). Invoking the hierarchy
$m_L \ll m_R \ll M$, the mass eigenvalues have approximate
magnitudes $m_d \simeq m_L m_R/M$ and $m_D \simeq M$. 
The ratio $m_R/M$ is
a hierarchy parameter, suppressing the down-quark charge
relative to the electroweak scale. The light
eigenstate has a predominant $d$ admixture, while the heavy
eigenstate is mostly the exotic $D$. The other charged fermions
develop similar mass matrices. (The electroweak scale mass
for the top quark demands that it, at least, not be see-saw
suppressed. I will come back to this point later, 
during a digression.)

The neutrino mass matrix has a very interesting form:
\begin{equation}
\left( \begin{array}{cccc} \overline{\nu}_L & \overline{(\nu_R)^c}
& \overline{N}_L & \overline{(N_R)^c}
\end{array} \right)
\left( \begin{array}{cccc} 
0 & 0 & m_L & m'_L \\ 
0 & 0 & m_R & m'_R \\
m_L & m_R & M_{LL} & M_{LR} \\
m'_L & m'_R & M_{LR} & M_{RR}
\end{array} \right)
\left( \begin{array}{c} 
(\nu_L)^c \\ \nu_R \\ (N_L)^c \\ N_R
\end{array} \right) + H.c.
\end{equation}
The eigenvalues have approximate magnitudes
\begin{equation}
\frac{m_L^2}{M} \ll m_{u,d,e},\quad
\frac{m_R^2}{M} \gg m_{u,d,e},\quad M,\quad M,
\end{equation}
where $M$ denotes a generic large scale set by the
parameters $M_{LL,LR,RR}$. The lightest eigenstate
is the $\nu_L$ up to small admixtures. We see that
the active neutrino mass is {\it automatically} more
suppressed (by a factor $m_L/m_R$) than those of the charged
fermions! This is the beauty of the universal see-saw
mechanism.

But the point I will emphasise here is that the spectrum
of heavy sterile neutrinos is quite interesting. There
are two extremely heavy fully sterile states of mass
set by the largest scale $M$. But there is also a heavy
weakly-sterile state of mass $m_R^2/M$ which feels the
right-handed weak interactions. It might be interesting
to analyse the leptogenesis implications of this
kind of heavy sterile spectrum, and to compare the
results to those derived from the standard see-saw model.

I come now to my third and final case study: the mirror
see-saw mechanism \cite{mirrorseesaw}. 
We return to the $G_{SM} \otimes G_{SM}$
model introduced above. In addition to the standard
left-handed neutrino $\nu_L$ and its mirror partner $\nu'_R$,
we introduce singlet states comprising the ``standard''
right-handed neutrino $\nu_R$ and its mirror partner $\nu'_L$.
The discrete symmetry between the sectors includes the
interchanges
\begin{equation}
\nu_L \leftrightarrow \nu'_R,\qquad 
\nu_R \leftrightarrow \nu'_L,
\end{equation}
where I have suppressed the Lorentz structure required
for a parity transformation.

The neutrino mass matrix is given by
\begin{equation}
\left( \begin{array}{cccc} \overline{\nu}_L & \overline{(\nu'_R)^c}
& \overline{(\nu_R)^c} & \overline{\nu}'_L
\end{array} \right)
\left( \begin{array}{cccc} 
0 & 0 & m_1 & m_2 \\ 
0 & 0 & m_2 & m_1 \\
m_1 & m_2 & M_1 & M_2 \\
m_2 & m_1 & M_2 & M_1
\end{array} \right)
\left( \begin{array}{c} 
(\nu_L)^c \\ \nu'_R \\ \nu_R \\ (\nu'_L)^c
\end{array} \right) + H.c.
\end{equation}
Superficially this looks similar to the universal see-saw
matrix, but it is actually very different. The entries
$m_{1,2}$ are both of the electroweak scale, while $M_{1,2}$
are large bare masses within the gauge singlet sector.
Notice that $m_1$ is proportional to the VEV of
the standard Higgs doublet, whereas $m_2$ is driven by the
VEV of the mirror Higgs doublet. In the version of the model
I am discussing, these two VEVs are equal: the discrete parity
symmetry is {\it not} spontaneously broken, with improper
Lorentz transformations consequently being exact symmetries
of both the Lagrangian and the world.

The consequences of this matrix are best revealed by
changing to the parity eigenstate basis,
\begin{equation}
\nu^{\pm}_L \equiv \frac{\nu_L \pm (\nu'_R)^c}{\sqrt{2}},\quad
\nu^{\pm}_R \equiv \frac{\nu_R \pm (\nu'_L)^c}{\sqrt{2}}
\end{equation}
which furnishes the mass matrix
\begin{equation}
\left( \begin{array}{cccc}
0 & 0 & 0 & m_+ \\
0 & 0 & m_- & 0 \\
0 & m_- & M_- & 0 \\
m_+ & 0 & 0 & M_+
\end{array} \right).
\end{equation}
The two $2\times 2$ blocks make it easy to see that the 
eigenvalues are of order
\begin{equation}
\frac{m_+^2}{M_+},\quad \frac{m_-^2}{M_-},\quad M_+,\quad
M_-,
\end{equation}
with eigenstates that are predominantly
$\nu^+_L$, $\nu^-_L$, $\nu^+_R$ and $\nu^-_R$, respectively.

The two light eigenstates therefore form a maximally-mixed
active-mirror pair, with the mirror component being the
weakly sterile $\nu'_R$. This is reminiscent of pseudo-Dirac
structure, but different from it. The important thing is
that the mirror see-saw mechanism automatically provides
light effectively-sterile neutrinos. 
The two heavy eigenstates are a pair
of maximally mixed fully-sterile states. Some cosmological
consequences of the heavy states have been explored in Ref.\cite{nfb}.

\subsubsection{Heavy active neutrinos}

We have seen that most candidates for heavy sterile neutrinos
actually provide weakly sterile varieties which, of course,
are not actually sterile at all. The active/sterile distinction
is thus arguably less useful for heavy states. Nevertheless
for completeness I will now review particle physics motivations
for heavy neutral leptons that feel the standard left-handed
weak interaction.

Three classes immediately spring to mind: a fourth standard
family, but with a neutrino more massive than 45 GeV; an
exotic family composed of higher weak-isospin representations
that contain neutral leptons; and vector-like weak isospin
doublet leptons.

A fourth family with a heavy neutrino is a possibility, but
there does not appear to be much motivation for it, and its
properties would in any case be quite constrained by precision
electroweak phenomenology. I will consider this scenario no
further. Exotic families are fun to think 
about \cite{exoticgens}, but, again,
there does not at present seem to be sufficient motivation to
spend much time on them. Apart from noting that
compositeness and weak-isospin-cubed models provide some
theoretical underpinning, we will move on.

Vector-like weak isospin doublet leptons have the
SU(3)$_c\otimes$SU(2)$_L\otimes$U(1)$_Y$ quantum
numbers
\begin{equation}
\psi_L \sim (1,2)(-1),\qquad \psi_R \sim (1,2)(-1),
\end{equation}
so the Dirac mass term $M \overline{\psi}_L \psi_R + H.c.$
is invariant under $G_{SM}$. Such particles can arise in
extended theories where the $\psi$'s transform chirally
under a higher symmetry that spontaneously breaks (eventually)
to $G_{SM}$.
The heavy neutral leptons are the $\tau_{3L}=+1$ components
of $\psi$.

As an interesting example, consider the subgroup chain
\begin{equation}
{\rm E}_6 \to {\rm SO}(10) \to {\rm SU}(5) 
\to {\rm SU}(3)_c \otimes {\rm SU}(2)_L \otimes {\rm U}(1)_Y.
\end{equation}
The smallest nontrivial representatation of E$_6$ decomposes
as:
\begin{equation}
27 \to 16 \oplus 10 \oplus 1 \to (5^* \oplus 10 \oplus 1) \oplus
(5^* \oplus 5) \oplus 1.
\end{equation}
The $16$ of SO(10) contains a standard family, while the $10$
of SO(10) contains a vector-like pair of lepton doublets (and
charge $-1/3$ quarks). These SO(10) ten-plet fermions will pick up
masses when E$_6$ breaks to SO(10), because they are chiral
under the former but vector-like under the latter (the $27$
of E$_6$ is a complex representation, whereas the $10$
of SO(10) is real).

I do not know if heavy active neutrinos such as these have
important cosmological applications. However, I would like to
digress about an interesting role the exotic fermions provided by
the $27$ of E$_6$ could play in the flavour problem.

We have seen that the $27$ of E$_6$ can supply heavy vector-like
states to partner down quarks and charged leptons. 
It also contains four neutral leptons in addition to the standard neutrino.
It is amusing that exotic partners for up quarks are absent by
group theoretic necessity. The raw ingredients therefore exist to
explain why 
\begin{equation}
m_t \sim 200\ {\rm GeV} \gg m_{b,\tau} \gg m_{\nu}.
\end{equation}
One invokes a not-quite-universal see-saw mechanism whereby
the bottom and tau lepton masses are suppressed relative 
to the electroweak scale through mixing with their exotic
partners, while the top quark, lacking a partner, cannot have
a suppressed mass. One would expect to be able to doubly suppress
the lightest active neutrino. This idea works for the third family;
obviously some other physics must be invoked to explain why
the charm and up masses are much less than the electroweak scale.

Davidson and I invented this framework in 1999 \cite{dv}; 
Rosner independently 
proposed similar ideas at about the same time \cite{rosner}. 
While a complete model
exploiting the above vision is still lacking, especially in the
neutrino sector, it might be useful to sketch some work-in-progress.

The Higgs multiplets necessary for fermion mass generation can
also be found in the 27 of E$_6$. Different components of the
Higgs multiplet (or multiplets) must pick up hierarchical VEVs
to implement the mechanism. The bottom and tau mass matrices
are of the form
\begin{equation}
\left( \begin{array}{cc}
0 & \ell \\ x & M
\end{array} \right),
\end{equation}
where $\ell$ is an electroweak scale mass and $x \ll M$ are given by 
high symmetry breaking scales, with $x/M$ being the intra-family
hierarchy parameter.

We have explored three incarnations that provide different choices
for which symmetries are broken at scales $x$ and $M$. The three
versions utilise the three different electric charge embeddings
existing within E$_6$: standard, flipped and double-flipped.
To understand this feature, we need to return to the subgroups
of E$_6$ and look at them in more detail. A chain of maximal
subgroups is
\begin{eqnarray}
{\rm E}_6 \to {\rm SO}(10) \otimes {\rm U}(1)'' & \to &
{\rm SU}(5) \otimes {\rm U}(1)' \otimes {\rm U}(1)'' \nonumber\\
& \to & {\rm SU}(3) \otimes {\rm SU}(2) \otimes 
{\rm U}(1)_{Y_{\rm st}} \otimes {\rm U}(1)' \otimes {\rm U}(1)''.
\end{eqnarray}
The standard hypercharge (and hence electric charge) embedding
takes $Y = Y_{\rm st}$; the flipped embedding sees $Y$ as
a certain linear combination of $Y_{\rm st}$ and the generator
of U(1)$'$; the double-flipped case requires also an admixture
of the generator of U(1)$''$. It is fairly easy to explicitly
write down these three linear combinations, but I will refrain
from doing so here. 

The effect of flipping and double-flipping
is to rearrange the exotic vector-like fermions within alternate
SU(5) components of the $27$, hence affecting the mass generation
mechanism. For the flipped and double-flipped embeddings, the
highest scale $M$ is associated with the breaking of 
U(1)$'\otimes$U(1)$''$ down to a U(1) subgroup, while $x$
is correlated with the breakdown of SU(5) times this leftover
U(1) down to the standard model group $G_{SM}$. For the
standard embedding, $M$ and $x$ are related to a two-step
breakdown of U(1)$'\otimes$U(1)$''$ to a U(1) subgroup
and then to nothing, with SU(5) remaining exact.

\subsection{Oscillation effects}

Neutrino disappearence and flavour transformations are now very
well established experimentally \cite{nuoscdata}. 
Standard mass and mixing driven
oscillations are most probably responsible for these effects.
It is therefore important to explore the implications of
oscillations for cosmology.\footnote{The cosmological effects of
absolute neutrino masses constitute a complementary area of study
that I will not review here.} 

Oscillations between light-active and light-sterile neutrino
flavours can give rise to dramatic effects for Big Bang
Nucleosynthesis (BBN), the topic of the next section. 
Elsewhere in these proceedings Y. Wong describes how transformations
amongst the active neutrinos themselves can lead to neutrino asymmetry
equilibration, an important issue for BBN \cite{equil}. 
I will focus instead
on how heavy-heavy neutrino oscillations can be responsible for the 
baryon asymmetry of the universe through a leptogenesis 
mechanism.\footnote{Light-heavy oscillations are quickly decohered
by collisions with the background plasma.}

As we have seen, heavy sterile neutrinos are a generic consequence
of see-saw mechanisms. In the leptogenesis scenario of Fukugita
and Yanagida, out-of-equilibrium decays of heavy neutral leptons
into Higgs bosons and doublet neutrinos are used to generate a
lepton asymmetry that is then reprocessed by non-perturbative
effects in the SM (sphaleron-induced transitions) into a baryon 
asymmetry \cite{fy}. 
P. di Bari discussed this mechanism at the Workshop.
I will review an alternative idea, due to Akhmedov, Rubakov and
Smirnov \cite{ars}.

As in the Fukugita-Yanagida model, three heavy neutral lepton
singlets $N_{A,B,C}$ are produced in the cosmological plasma
through Yukawa couplings to the standard doublet neutrinos
and Higgs bosons. The dominant process involves top quarks
because of their large Yukawa coupling constant.
The production is CP symmetric, so
equal numbers of $N$'s and $\overline{N}$'s are created.

The Yukawa coupling constants governing the $N$-Higgs-$\nu$
interactions are chosen in a mild hierarchy so that one
or two (but not all three) of the heavy singlets are
brought into thermal equilibrium prior to the electroweak
phase transition at a temperature 
$T_{EW} \simeq 100$ GeV. For definiteness,
we will follow the inventors by supposing that $N_A$ and $N_B$
are equilibrated by $T_{EW}$, but $N_C$ is not.

CP violation exists in the mixing matrix relating the
Higgs interaction $N$-eigenstates and the mass eigenstates.
After their production, the $N_{A,B}$ fermions begin
oscillating in a CP asymmetric way while continuing to
interact with the background plasma. The CP violating
mixings serve to separate the overall zero lepton
number into nonzero asymmetries $L_{A,B,C}$, maintaining
$L_A + L_B + L_C = 0$. The fast Yukawa interactions of
$N_A$ and $N_B$ transfer their asymmetries to standard
doublet neutrinos during the epoch prior to the
electroweak phase transition when sphaleron processes
are rapid. A nonzero baryon asymmetry results. 
Crucially, the $C$-type
asymmetry is never reprocessed into the baryon sector.
After the electroweak phase transition, it gets
transferred to the light neutrino sector, but by then sphaleron
transitions have switched off.

The parameter ranges required for this mechanism to work
are different from that of Fukugita-Yanagida. Relatively
small singlet masses, in the GeV to 10's of GeV range, are
acceptable for instance. Small Yukawa coupling constants
and mixing angles are also a feature according to Ref.\cite{ars}.

\section{Update on BBN and sterile neutrinos}

Big Bang Nucleosynthesis begins at about $T \simeq 0.8$ MeV,
when nuclear statistical equilibrium no longer obtains \cite{kt}.
Neutrinos feature in two important ways: by contributing
relativisitic energy density to the plasma thus helping to
drive the expansion of the universe, and through the
beta equilibrium processes
\begin{equation}
n\, \nu_e \leftrightarrow p\, e^-,\quad 
p\, \overline{\nu}_e \leftrightarrow n\, e^+,\quad
n \leftrightarrow p\, e^-\, \overline{\nu}_e,
\end{equation}
that maintain the neutron to proton ratio at
equilibrium values. Active to light-sterile oscillations can
increase the expansion rate because of sterile neutrino
production and alter beta-equilibrium
by affecting the electron-neutrino and antineutrino
distribution functions. The light elements produced through
BBN have abundances that depend on the neutron to proton
ratio and the relative sizes of nuclear reaction rates
and the expansion rate. The primordial abundances thus provide
a probe of active-sterile mixing (assuming that is the only
modification to standard BBN).

The paradigm known as ``standard BBN'' is based on the
following assumptions:
\begin{itemize}
\item There are three neutrino and antineutrino flavours and 
they are massless.
\item All six have Fermi-Dirac distributions with zero chemical
potentials.
\item There is one free parameter, the baryon to photon ratio $\eta$.
It also implicitly assumed that this parameter is homogeneous.
\end{itemize}
The outputs are the primordial abundances of $^4$He, D, $^3$He and
$^7$Li.
Before comparing standard BBN with observational data, I would like
to comment on the assumptions listed above. 

If terrestrial experiments establish the existence of 
significant active-sterile mixing, then the first
assumption of standard BBN, as quoted above, will have to be 
modified. We will see
very shortly what current measurements say about the relativistic
energy density during BBN.

For temperatures above the neutrino decoupling temperature of
about 1 MeV, Fermi-Dirac (FD) distributions
are very well justified because standard weak interactions
keep all active neutrino flavours in kinetic equilibrium. 
After decoupling, the FD forms are maintained unless 
oscillation effects are operative.

Chemical equilibrium is maintained between neutrinos and
antineutrinos of each active flavour $\alpha = e,\mu,\tau$
by well-established electroweak processes
such as $\nu_\alpha\, \overline{\nu}_{\alpha} 
\leftrightarrow e^+\, e^- \leftrightarrow \gamma\, \gamma$.
This enforces the relation $\mu_{\nu_{\alpha}} =
- \mu_{\overline{\nu}_{\alpha}}$ between the chemical potentials
above the $\alpha$-flavour chemical decoupling temperature.
Setting these chemical potentials to zero is an extra
assumption adopted within standard BBN. 
The conditions $\mu_{\nu_e} = \mu_{\nu_\mu} = \mu_{\nu_\tau}$
and the analogous one for antineutrinos, of which the
standard BBN zero chemical potential assumption is a special
case, will be
maintained under pure active-active oscillations, but
are in general not dynamically consistent if active-sterile 
oscillations occur. The most dramatic manifestation of
this is the exponentially fast creation of neutrino-antineutrino
asymmetries through collision-affected active-sterile oscillations
when certain conditions are satisfied \cite{nuasym}.\footnote{Nonzero
chemical potentials imply unequal number densities for neutrinos
and antineutrinos.}

The zero chemical potential assumption is adopted partly for
simplicity and partly because the small observed baryon asymmetry,
$\eta \sim 10^{-10}$, encourages one to suppose the neutrino
asymmetries have similar magnitudes. However, because relic
neutrinos have not been directly detected, the observational
constraints on their asymmetries are quite weak. At least two
mechanisms for producing large neutrino asymmetries have been
explored -- through active-sterile oscillations (as mentioned above)
and Affleck-Dine processes \cite{gelmini} -- so there is no requirement
for the neutrino asymmetries to be as small as their baryon counterpart.

Through the matter effect \cite{msw}, nonzero asymmetries can suppress sterile
neutrino production \cite{prl}. This must be taken into account when computing
BBN relativistic energy density constraints on active-sterile mixing.
An $e$-like asymmetry is a special case of distribution function
modifications that affect beta-equilibrium and hence the $^4$He
abundance prediction.

Finally, standard BBN for simplicity assumes spatial homogeneity.
As primordial abundance determinations improve, it will be
interesting to see if reality reflects this simplicity. 
One interesting possibility in the inhomogeneous case is for
a spatially varying $e$-like asymmetry to be generated by
active-sterile oscillations, seeded by an inhomogeneous baryon
distribution \cite{inhombbn}.

\begin{figure}[htbp]
  \includegraphics[height=.5\textheight]{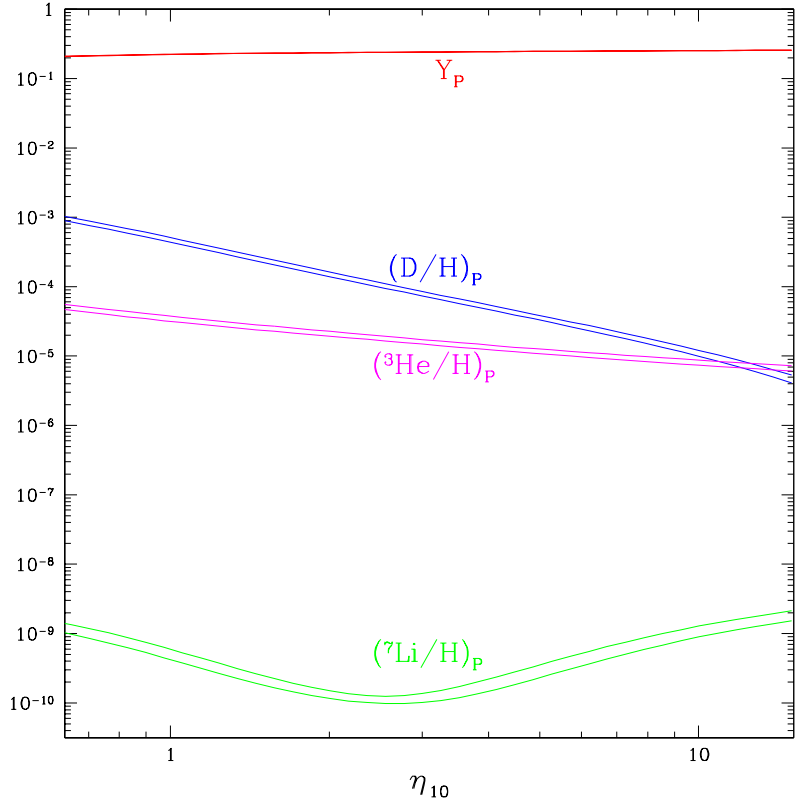}
  \caption{Predictions of standard BBN as a function of the
baryon-to-photon ratio $\eta_{10} \equiv \eta \times 10^{10}$.
Figure taken from Ref.\cite{gary} and reproduced by permission
of Cambridge University Press.
}
\end{figure}

The prediction of standard BBN is summarised in Figure 1, taken
from the recent review by G. Steigman \cite{gary}. One can see that $^4$He
is by far the most abundant species, and that it is quite
insensitive to $\eta$. For phenomenologically acceptable $\eta$'s,
deuterium is the next most common isotope, falling fairly steeply
with the baryon density (because it is easily destroyed). I will
focus on these two species. Current measurements are summarised
in Figs.\ 2-4, taken once again from the Steigman
review \cite{gary}. One can see that the $^4$He data clearly 
approach an asymptotic value with decreasing metallicity, as
tracked by oxygen in this case. The deuterium data show
more scatter, both as a function of redshift and metallicity.

\begin{figure}[htbp]
  \includegraphics[height=.5\textheight]{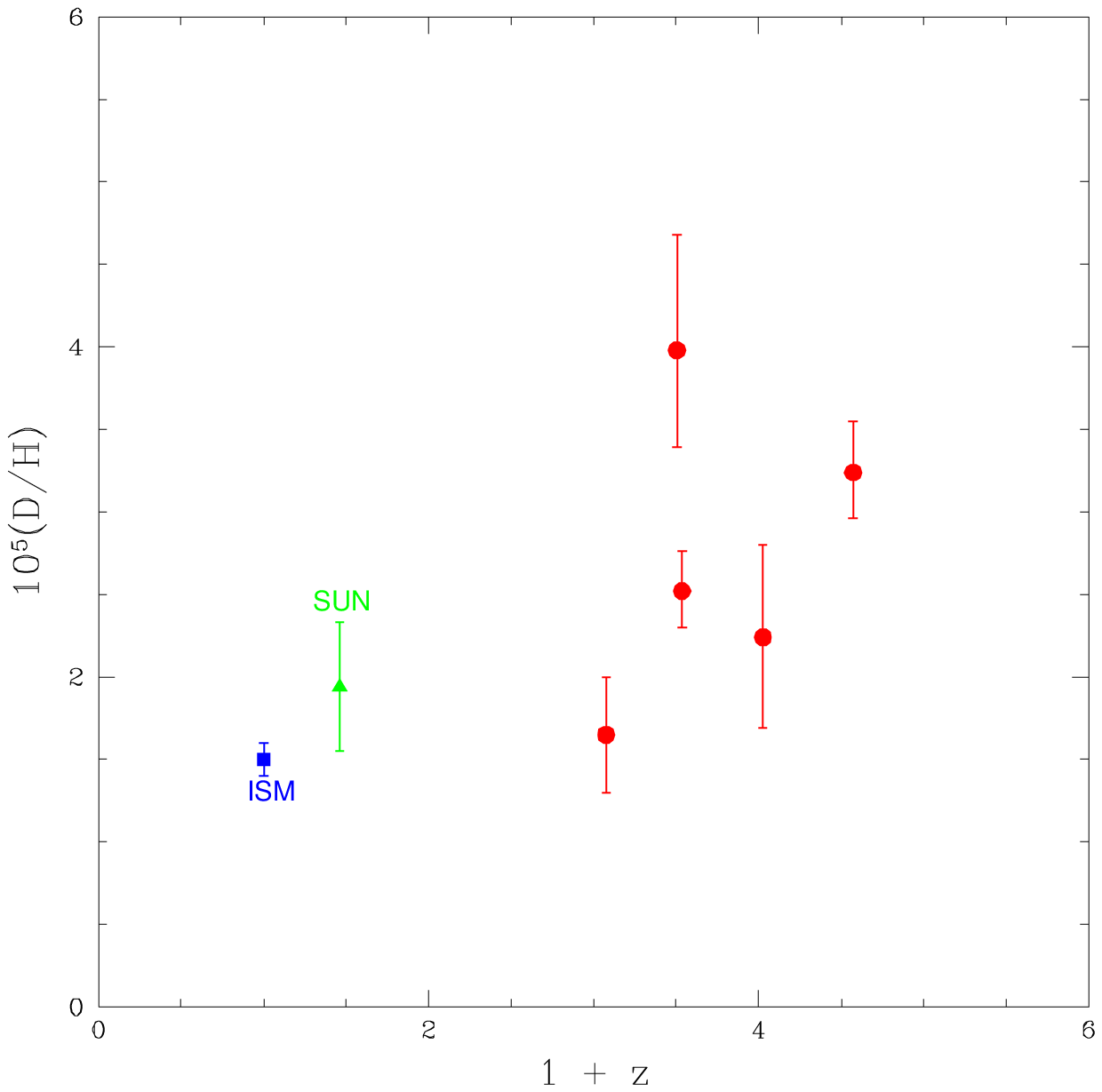}
  \caption{Deuterium abundance as a function of red-shift. The deuterium
fractions of the sun and the interstellar medium (ISM) are given for 
comparison purposes. Figure taken
from Ref.\cite{gary} and reproduced by permission of Cambridge University Press.
}
\end{figure}

\begin{figure}[htbp]
  \includegraphics[height=.5\textheight]{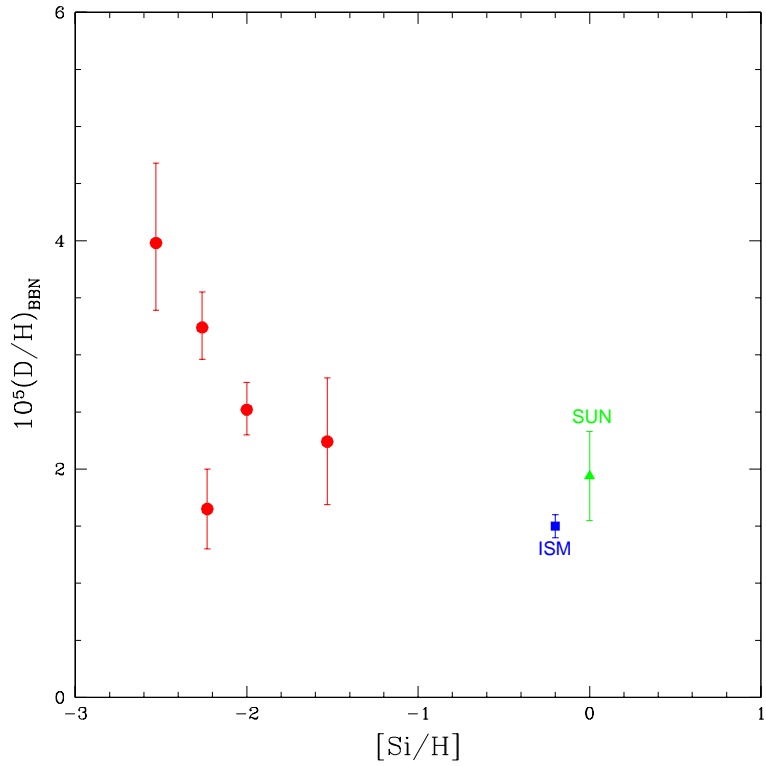}
  \caption{Deuterium abundance as a function of metallicity. Figure
taken from Ref.\cite{gary} and reproduced by permission from Cambridge
University Press.
}
\end{figure}

\begin{figure}[htbp]
  \includegraphics[height=.3\textheight]{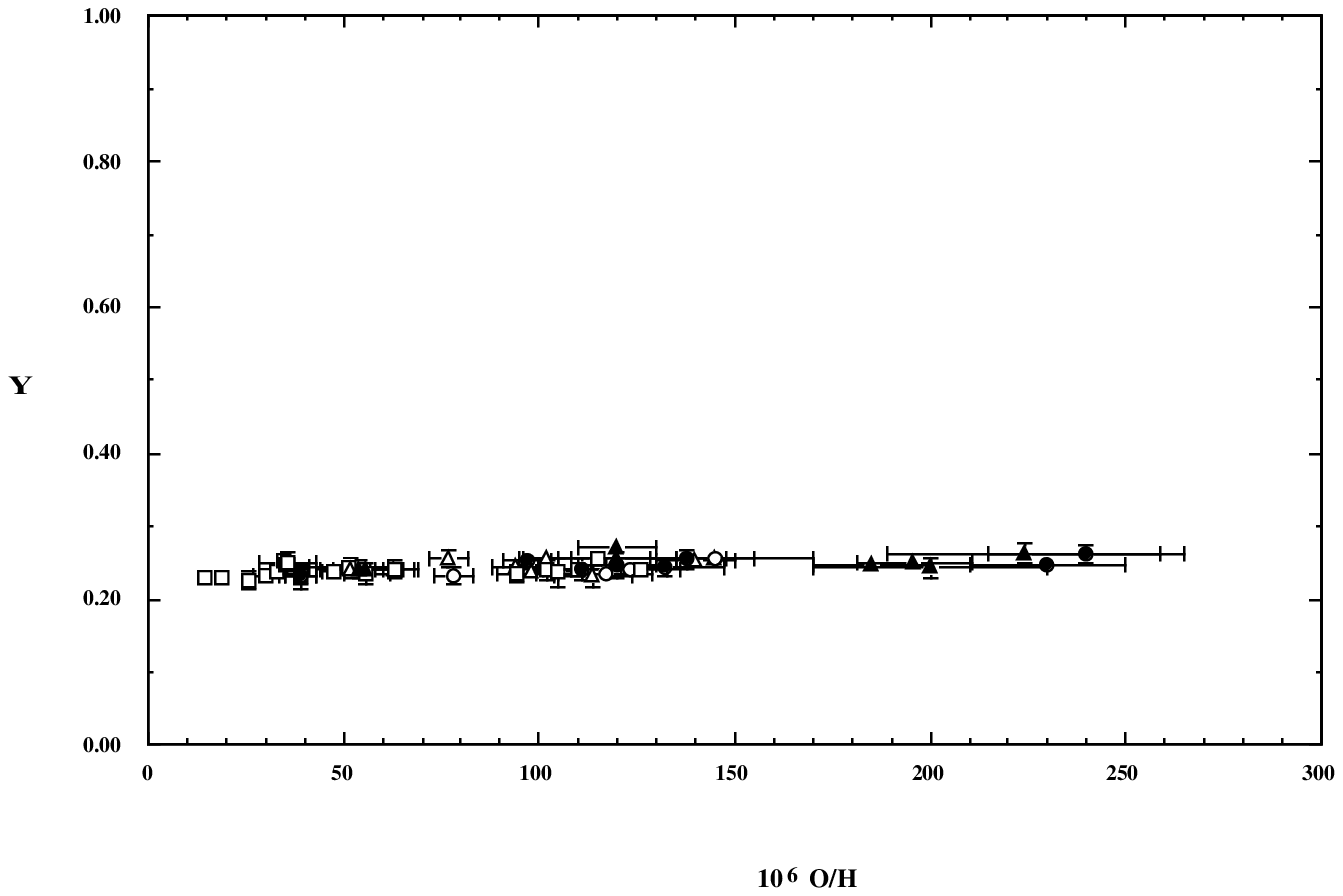}
  \caption{Helium abundance (mass fraction) as a function of metallicity.
Figure taken from Ref.\cite{gary} and reproduced by permission of
Cambridge University Press.
}
\end{figure}

The implications of these measurements have been recently
summarised by di Bari \cite{pasq}, and even more recently by Steigman \cite{gary}. 
At the time of the di Bari paper, there were two basically
incompatible $^4$He abundance determinations, the so-called
``high value''
\begin{equation}
Y_p = 0.244 \pm 0.002
\end{equation}
of Izotov and Thuan (IT) \cite{it}, and the ``low value''
\begin{equation}
Y_p = 0.234 \pm 0.003
\end{equation}
of Olive and Steigman (OS) \cite{os}. The treatment of systematic
effects accounts for the quite serious disparity.
The dispassionate may choose to explore the ramifications
of these extractions separately, or adopt a ``compromise''
value, such as the
\begin{equation}
Y_p = 0.238 \pm 0.005
\end{equation}
figure suggested by OS which inflates the error
to reflect the systematic uncertainties. I prefer to
treat the two determinations separately.

The deuterium abundance range advocated by O'Meara et al \cite{omeara} is
\begin{equation}
\frac{D}{H} = 3.0 \pm 0.4.
\label{omeara}
\end{equation}
After considering all of the available data,
Steigman argues for a larger systematic uncertainty
and suggests the range
\begin{equation}
\frac{D}{H} = 3.0^{+1.0}_{-0.5}
\end{equation}
as a safer alternative.

The internal test of standard BBN is whether a unique value for $\eta$
fits both the $^4$He and deuterium data. The inferred $3\sigma$ baryon
density ranges are:
\begin{eqnarray}
\eta({\rm High\ Y}) & = & 2.0 - 7.0,\nonumber\\
\eta({\rm Low\ Y}) & = & 0.6 - 3.5,\nonumber\\
\eta(D/H) & = & 4.5 - 7.7,
\end{eqnarray}
where the deuterium range of Eq.\ (\ref{omeara}) has been
used. (The semi-analytical approach of di Bari \cite{pasq} is a very convenient
tool when faced with extracting $\eta$ from many
different primordial abundance determinations.) One can see
that the IT and D/H data yield a consistent $\eta$, whereas
OS and D/H are inconsistent at a greater than $3\sigma$ level.

The recent cosmic microwave background radiation (CMBR) anisotropy 
data now provide an important {\it independent} determination
of the baryon density \cite{cmbr}. A comparison of the CMBR value with that (or those!)
determined from primordial abundance considerations
therefore provides a very interesting external
test of standard BBN. The CMBR $3\sigma$ range is
\begin{equation}
\eta({\rm CMBR}) = 3.6 - 9.3.
\end{equation}
This is consistent with IT and D/H, but not with OS!

Clearly, the issue of systematic effects in the extraction of
the primordial $^4$He abundance is critical. Interestingly,
a reanalysis of the data used by IT to derive the High Y
range has recently been performed by Peimbert, Peimbert
and Luridiana (PPL) \cite{ppl} (see the discussion by Steigman \cite{gary}).
PPL take into account systematic effects involving the
temperature of different chemical components of the
clouds which supply the helium data. They conclude
that these effects lead to a {\it lower} $Y_p$ than
derived by IT! By ignoring another systematic effect
involving the collisional excitation of hydrogen, they
deduce
\begin{equation}
Y_p(PPL1) = 0.2356 \pm 0.0020,
\end{equation}
whereas a preliminary result incorporating the collisional
effect also is
\begin{equation}
Y_p(PPL2) = 0.2384 \pm 0.0025,
\end{equation}
which again raises the figure slightly.

My summary of the situation is in Fig.\ 5. As you can see,
the PPL reanalysis apparently strengthens the evidence
that standard BBN is not completely consistent with the
best elemental abundance and CMBR analyses we have at present.
In a nutshell, the CMBR and D/H data are consistent with
each other, but are inconsistent at a reasonably significant
level with the helium data. An overlap is obtained only when
extremes of the $3\sigma$ ranges are used. Given the reasonable
doubts one may have about how well systematic effects are
understood, I would not want to draw too strong a conclusion
just yet. However, it is fair to say that there is a tension
between the data sets quoted above. PPL are promising to
update their preliminary $Y_p$ figure with the collisional
effect incorporated. If it should move up, even a little,
relative to what I have called PPL2, then the $\eta$ range
will move towards consistency with deuterium and CMBR (recall
that small changes in $Y_p$ lead to large changes in the
inferred $\eta$). Perhaps the most dramatic new information
we expect in the near future is a more precise CMBR figure
for $\eta$. The MAP experiment promises to reduce the
uncertainty to about $10\%$, while the future Planck
mission may reduce it all the way to about $1\%$. It is
easy to see from Fig.\ 5 that a high and precise CMBR $\eta$ could
really create a crisis for standard BBN.

\begin{figure}
  \includegraphics[height=.3\textheight]{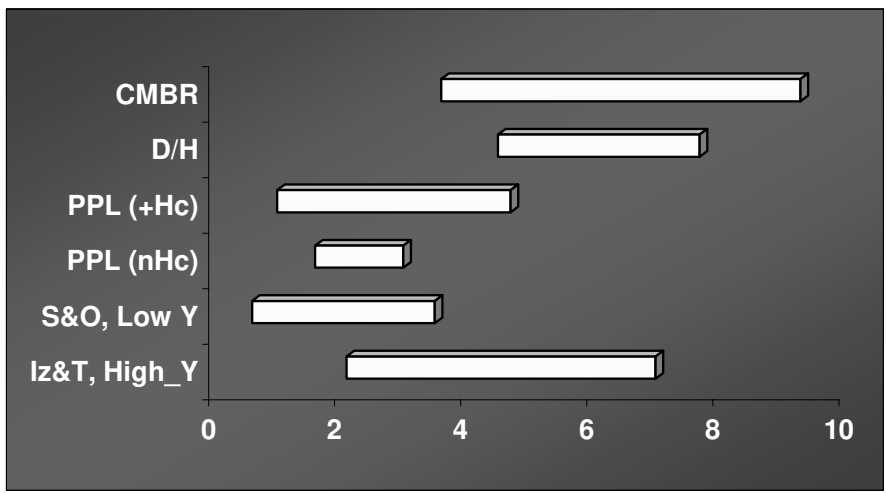}
  \caption{Summary of allowed $3\sigma$ ranges for the baryon-to-photon
ratio $\eta_{10} \equiv \eta \times 10^{10}$ (horizontal axis) 
from different observations. Standard BBN would be successful were
a single value of $\eta$ consistent with all the data. 
The different data sets are: CMBR (cosmic microwave background acoustic peaks), 
$D/H$ (primordial deuterium abundance), PPL+Hc (primordial Helium
abundance from Peimbert et al with preliminary incorporation of collision
effect), PPLnHc (same with collision effect correction removed),
SO (Steigman and Olive ``low-Y result'' for Helium), Iz\&T (Izotov
and Thuan ``high-Y result'').
}
\end{figure}

This rather interesting situation certainly provides
some motivation for thinking about non-standard BBN.
One way to dramatise the situation is to point out
that a lower helium yield can be obtained by reducing
the expansion rate of the universe. Parameterising the
latter in terms of an effective number of neutrino flavours
in equilibrium during BBN, one can fit the light element
abundance data better by reducing this parameter from
three to about two and a half. Of course, no known physics can
actually prevent the full equilibration of all
three active species. However, there is a well-known
modification of standard BBN that can reduce the helium yield
without affecting the expansion rate appreciably: the
introduction of a nonzero chemical potential for
electron neutrinos and antineutrinos \cite{kneller}. Using the
equilibrium value for the neutron to proton
ratio
\begin{equation}
\frac{n_n}{n_p} = \exp\left[ \frac{m_p - m_n}{T} 
- \frac{\mu_{\nu_e}}{T} \right],
\end{equation}
we can estimate that $\xi_e \equiv \mu_{\nu_e}/T
\sim 0.05$ will reduce $Y_p$ by about $0.01$, and
thus provide a good simultaneous fit to the helium,
deuterium and CMBR data.

Interestingly, active-sterile neutrino oscillations
can induce significant chemical potentials or asymmetries \cite{nuasym}.
So it has been suggested that rather than being a cosmological
liability, light sterile neutrinos might actually be
required for a fully successful BBN, with various models
explored in recent years \cite{xie}. (These scenarios may involve more
than just nonzero chemical potentials -- 
oscillation-induced spectral distortion away from Fermi-Dirac form
for $T < 1$ MeV may also occur.) 
However, new experimental data,
especially from SNO, and a much better understanding of
active-active oscillations in the early universe \cite{equil} now 
require all of these scenarios to be re-analysed. 

The active-active story is a rather interesting one.
Neutrino oscillation dynamics in the early
universe is intrinsically non-linear because neutrinos
scatter off other neutrinos in the plasma. For the 
active-active case, it turns out that this tends to
induce synchronisation \cite{synch}: neutrinos of all energies
are driven to oscillate at the same rate, quite 
unlike the behaviour when neutrino-neutrino scattering
is absent. Perhaps even more importantly, neutrino
asymmetries in this case do {\it not} suppress
oscillation amplitudes, unlike the better-studied active-sterile
situation. Yvonne Wong will explain this more 
fully in her talk. If an active-active 
$\nu_{\alpha} \leftrightarrow \nu_{\beta}$ channel is governed
by a large mixing angle, then the $\alpha-$ and $\beta-$like
asymmetries will tend towards equilibration. If both the
solar and atmospheric neutrino deficits are due to
large-angle active-active mixing, and if the solar
$\Delta m^2$ is at the upper end of the allowed range,
about $10^{-5}$ eV$^2$, then all the asymmetries 
equilibrate. In these circumstances, the BBN constraints
on $L_{\nu_e}$ apply to $L_{\nu_{\mu,\tau}}$ as well \cite{equil}.
These developments are still new: no one has yet calculated
neutrino distribution function and asymmetry evolution
in a light sterile neutrino model with active-active
transitions correctly accounted for.

Let us now turn to BBN bounds on the relativistic energy
density and distribution function distortion: when would
agreement between calculations and observation be unambiguously
absent? Introduce the parameters $\Delta N_{\nu}^{\rho}$
and $\Delta N_{\nu}^{f}$, where the former parameterises
the change in the relativistic energy density in terms
of an effective change in the neutrino flavour count, while
the latter is a coarse-grained measure of the effect of spectral
alteration away from zero chemical potential FD form.
Following di Bari \cite{pasq}, we proceed by writing down approximate
analytical fits (valid around $\eta \simeq \eta_{CMB}$)
to the primordial abundance yields as a
function of $\Delta N_{\nu}^{\rho}$:
\begin{eqnarray}
Y_p(\eta,\Delta N_{\nu}^{\rho}) & \simeq & Y_p^{SBBN}(\eta)
+ 0.0137 \Delta N_{\nu}^{\rho},\nonumber \\
\left( \frac{D}{H} \right)(\eta, \Delta N_{\nu}^{\rho})
& \simeq & \left( \frac{D}{H} \right)^{SBBN}(\eta)
[1 + 0.135 \Delta N_{\nu}^{\rho}]^{0.8}
\end{eqnarray}
Spectral distortion changes the helium yield but negligibly
affects deuterium. The course-grained spectral distortion
parameter may thus be defined by
\begin{equation}
\Delta N_{\nu}^{f} \equiv \frac{Y_p(\eta,\Delta N_{\nu}^{\rho}, \delta f)
- Y_p(\eta,\Delta N_{\nu}^{\rho})}{0.0137}
\end{equation}
where $\delta f$ denotes the deviation of the actual $\nu_e$ and
$\overline{\nu}_e$ distribution functions from zero chemical potential
FD form and is in general a complicated function.

By comparing the measured yields with $Y_p(\eta_{CMB})$ one obtains
$3\sigma$ allowed ranges for 
\begin{equation}
N_{\nu}^{tot} \equiv 3 + \Delta N_{\nu}^{\rho} + \Delta N_{\nu}^{f}.
\end{equation}
The preference for an effective neutrino number of less than three is
clearly displayed in Fig.\ 6. Note that deuterium yields only
weak constraints. Cosmic microwave background data also at present
provide only weak constraints on $\Delta N_{\nu}^{\rho}$.
The maximum increase in $N_{\nu}^{tot}$ one can
tolerate before BBN fails unambiguously is about $0.3$, taking the
original IT high-$Y$ figures uncritically. Note in particular, that
a fully equilibrated sterile flavour corresponding to $N_{\nu}^{tot} = 4$
would cause BBN failure by a comfortable margin (assuming the absence of
a negative $\Delta N_{\nu}^{f}$ that compensates for $\Delta N_{\nu}^{\rho}$).

\begin{figure}[htbp]
  \includegraphics[height=.3\textheight]{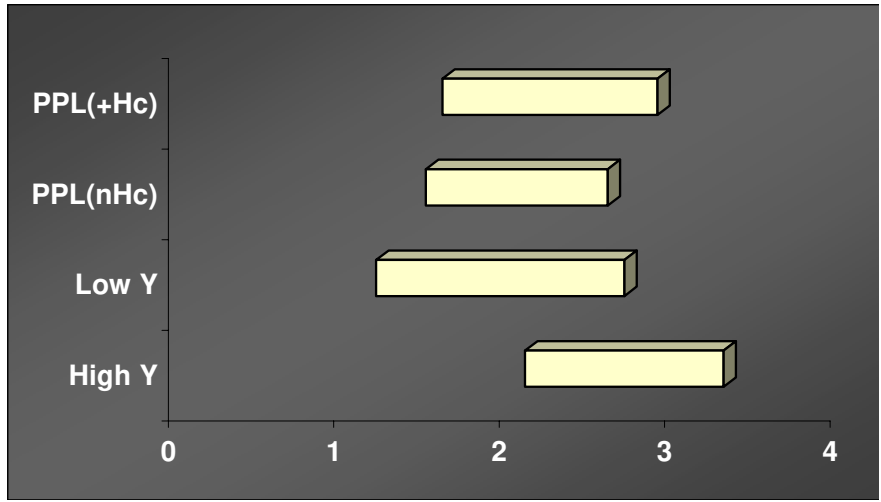}
  \caption{The $3\sigma$ ranges for the preferred effective total number of 
neutrino flavours implied by the different primordial helium abundance extractions.
See the caption of Fig.\ 5 for an explanation of the labels on the vertical axis.
}
\end{figure}

In the future, in addition to a baryon density at $10-1\%$ precision, CMBR
acoustic peak data may be precise enough to allow a probe of relativistic
energy density at the $\pm 0.1$ level as quantified by $\Delta N_{\nu}^{\rho}$.
If $\Delta N_{\nu}^{\rho} + \Delta N_{\nu}^{f} < 0$ continues to be supported
by helium data, especially if better understanding of systematic 
effects increases confidence in the extracted primordial abundance figure,
then a combination BBN and CMBR data could provide evidence for a
nonzero $e$-type asymmetry!\footnote{Beware that BBN and CMBR data probe
the universe during different epochs, so one has to assume that the
physics of the intervening period is understood when one draws conclusions
using combined information.}

As our final topic, we turn to the BBN cosmology of $2+2$ and $3+1$ 
neutrino models. This was analysed in great detail by 
di Bari \cite{pasq} (see also \cite{kev}), 
who concluded
that essentially
all of the possible scenarios yield parameters in the range
\begin{equation}
\Delta N_{\nu}^{\rho} = 0.9 - 1.0,\qquad \Delta N_{\nu}^{f} \simeq 0.
\end{equation}
Basically, the oscillation parameter conditions 
required for the active-sterile asymmetry
generation mechanism to work are not met, so negligible asymmetries
result and the sterile flavour is fully or almost fully equilibrated.
The only loophole is in the $2+2$ model which has the $\nu_{\mu,s}$
couplet lighter than the $\nu_{e,\tau}$ couplet. If the mixing angles
between $\nu_{e,\tau}$ and the lighter $\nu_s$ are very small, then
some asymmetry generation may occur, but not enough to prevent BBN
failure. In a couple of subcases, the addition of
a second sterile neutrino flavour can allow one to engineer an 
acceptable outcome, but the resulting model looks rather forced.

These conclusions motivate a reconsideration of the ``large pre-exisiting
neutrino asymmetry'' idea \cite{prl}. One supposes that some mechanism operating
at higher temperatures produces asymmetries that are large enough to
suppress active-sterile mixing via the matter effect and hence to 
also suppress sterile neutrino production. For instance, $\nu_{\mu}-\nu_s$
mixing with atmospheric range oscillation parameters requires 
an asymmetry high than about $10^{-5}$. While this is five
orders of magnitude higher than the baryon asymmetry, it is still very
small by BBN standards. In particular, an $L_{\nu_e} \sim 10^{-5}$ 
generated through the active-active asymmetry equilibration mechanism 
would have a negligible effect on the helium abundance. What happens
when one moves away from pure $\nu_{\mu}-\nu_s$ mixing is currently
under investigation.

A particularly amusing possibility is that
\begin{enumerate}
\item MiniBooNE confirms the LSND effect, with its indirect evidence for
a light sterile neutrino;
\item future long baseline experiments garner {\it direct evidence} for a
light sterile neutrino through neutral to charged current ratio 
measurements, thus confirming the oscillation explanation for
the LSND anomaly;
\item BBN remains incompatible with four (or more) equilibrated species.
\end{enumerate}
These circumstances would provide remarkable evidence for the existence
of fairly large pre-existing lepton asymmetries. This would obviously
be important information for high temperature models of leptogenesis
and baryogenesis.

\section{Conclusions}

The known neutrinos play an important role in cosmology, especially in
Big Bang Nucleosynthesis through the relativistic
energy density and hence expansion rate of the universe, and through
the neutron-proton interconversion reactions mediated by $\nu_e$ and
$\overline{\nu}_e$. Much more speculatively, heavy neutrino-like states
might be responsible for producing the cosmological baryon asymmetry
through the sphaleron reprocessing of a lepton asymmetry.

I have reviewed how additional neutrino-like states can arise in
extensions of the Standard Model of particle physics. Three see-saw
mechanisms -- the standard, the universal, the mirror -- were 
presented as case studies of how additional neutrino flavours
may be required for understanding mass generation and mass
hierarchies. All three furnish heavy neutrinos, and the mirror
see-saw supplies possibly the best candidate theory for light sterile
neutrinos.

The active/sterile distinction was critically reviewed, with sterility
subdivided into full-blown and weak varieties, depending on whether
or not those neutrino-like states couple to hypothetical gauge
interactions beyond the known left-handed weak interaction. The
additional interactions, if chiral, would provide a natural
mass scale to the weakly sterile neutrinos through spontaneous
symmetry breaking. Also, these interactions would affect the cosmology
of the heavy neutral leptons, and hence also mechanisms of
leptogenesis. I reviewed the Akhmedov-Rubakov-Smirnov leptogenesis
model as an alternative to the better known Fukugita-Yanagida
scenario.

I digressed on the subject of why the top quark has a weak scale
mass while the other third family fermions have much smaller
masses. I pointed out that the mathematics of the $27$ of $E_6$
through a not-quite-universal see-saw mechanism might provide
the answer.

The present status of Big Bang Nucleosynthesis was then discussed.
We observed that there is a tension between the helium abundance
data and both the deuterium and cosmic microwave background
data. A recent reanalysis of helium data by Peimbert, Peimbert
and Luridiana \cite{ppl} strengthens the case for the inconsistency of
standard BBN. The problem of systematic errors in the extraction
of primordial abundances cautions against overly strong conclusions,
however \cite{gary}. Future precise determinations of the baryon density
and relativistic energy density from CMBR acoustic peak
data promises to challenge standard BBN in important ways.

So, there is motivation to consider non-standard BBN. A nonzero
$e$-like asymmetry can resolve the tension by lowering the
helium abundance without affecting the deuterium abundance,
the latter being in good agreement with the independent 
CMBR determination of the baryon density. Such an asymmetry
can be generated by active-sterile oscillations. Previous
specific models of this type need to be revised in light
of the SNO solar neutrino results and our increased understanding
of the early universe dynamics of active-active oscillation channels.

Finally, we observed that the much discussed $2+2$ and $3+1$
models lead almost inevitably to equilibration of the sterile
flavour without the compensating generation of an $e$-like asymmetry.
This argues for the existence of asymmetries generated at higher
temperature scales of sufficient magnitude to suppress sterile
neutrino production (and avoid being washed out) \cite{prl}. Should MiniBooNE
confirm the LSND anomaly, and if primordial abundance data remain
inconsistent with four equilibrated neutrino species, then some
remarkable evidence will exist for significant neutrino asymmetries
generated at epochs significantly before that of BBN.

\begin{theacknowledgments}
I would like to thank the organisers for an enjoyable workshop and
for their support. 
This work was also supported by the University of Melbourne.
\end{theacknowledgments}

\end{document}